\documentclass[conference]{IEEEtran}

\usepackage{cite}

\usepackage[pdftex]{graphicx}
\DeclareGraphicsExtensions{.pdf,.jpeg,.png}

\usepackage{amsmath}
\usepackage{algorithmic}

\usepackage{array}

\usepackage{lipsum}

\newcommand\blfootnote[1]{%
  \begingroup
  \renewcommand\thefootnote{}\footnote{#1}%
  \addtocounter{footnote}{-1}%
  \endgroup
}

\usepackage{url}

\begin{document}
\title{Big Data Meets HPC Log Analytics: Scalable Approach to Understanding Systems at Extreme Scale}
\author{\IEEEauthorblockN{Byung H. Park\IEEEauthorrefmark{1}, 
			  Saurabh Hukerikar\IEEEauthorrefmark{1}, 
			  Ryan Adamson\IEEEauthorrefmark{2}, and 
			  Christian Engelmann\IEEEauthorrefmark{1}}
	\IEEEauthorblockA{\IEEEauthorrefmark{1}Computer Science and Mathematics Division\\
			  \IEEEauthorrefmark{2}National Center for Computational Sciences\\
		  	  Oak Ridge National Laboratory\\
		  	  Oak Ridge, TN, USA\\
			  Email: \{parkbh, hukerikarsr, adamsonrm, engelmannc\}@ornl.gov}
       }

\maketitle

\begin{abstract}
Today's high-performance computing (HPC) systems are heavily
instrumented, generating logs containing information about abnormal
events, such as critical conditions, faults, errors and failures,
system resource utilization, and about the resource usage of user
applications. These logs, once fully analyzed and correlated, can
produce detailed information about the system health, root causes of
failures, and analyze an application's interactions with the system,
providing valuable insights to domain scientists and system
administrators. However, processing HPC logs requires a deep
understanding of hardware and software components at multiple layers
of the system stack. Moreover, most log data is unstructured and
voluminous, making it more difficult for system users and administrators 
to manually inspect the data. With rapid increases in the scale and complexity of
HPC systems, log data processing is becoming a big data challenge.
This paper introduces a HPC log data analytics framework that is based
on a distributed NoSQL database technology, which provides scalability
and high availability, and the Apache Spark framework for rapid in-memory
processing of the log data. The analytics framework enables the extraction 
of a range of information about the system so that system administrators
and end users alike can obtain necessary insights for their specific
needs. We describe our experience with using this framework to glean
insights from the log data about system behavior from the Titan 
supercomputer at the Oak Ridge National Laboratory.
\end{abstract}


\blfootnote{This manuscript has been authored by UT-Battelle, LLC
  under Contract No. DE-AC05-00OR22725 with the U.S. Department of
  Energy. The United States Government retains and the publisher, by
  accepting the article for publication, acknowledges that the United
  States Government retains a non-exclusive, paid-up, irrevocable,
  worldwide license to publish or reproduce the published form of this
  manuscript, or allow others to do so, for United States Government
  purposes. The Department of Energy will provide public access to
  these results of federally sponsored research in accordance with the
  DOE Public Access Plan
  (http://energy.gov/downloads/doe-public-access-plan).}

\IEEEpeerreviewmaketitle

\section{Introduction}
\label{sec:Introduction}

Log data is essential for understanding the behavior of high-performance
computing (HPC) systems by recording their usage and troubleshooting
system faults. Today's HPC systems are heavily instrumented at every layer 
for health monitoring by collecting with performance counters and resource 
usage data. Most components also report information about abnormal 
events, such as critical conditions, faults, errors and failures. This 
system activity and event information is logged for monitoring 
and analysis. Large-scale HPC installations produce various types of log 
data. For example, job logs maintain a history of application runs, the 
allocated resources, their sizes, user information, and exit statuses, 
i.e., successful vs. failed. Reliability, availability and serviceability 
(RAS) system logs derive data from various hardware and software sensors, 
such as temperature sensors, memory errors and processor utilization. 
Network systems collect data about network link bandwidth, congestion and 
routing and link faults. Input/output (I/O) and storage systems produce 
logs that record performance characteristics as well as data about 
degradations and errors detected.

HPC log data, when thoroughly investigated both in spatial and
temporal dimensions, can be used to detect occurrences of failures and
 understand their root causes, identify persistent temporal and spatial
patterns of failures, track error propagation, evaluate system
reliability characteristics, and even analyze contention for shared
resources in the system. However, HPC log data is derived from multiple
monitoring frameworks and sensors and is inherently unstructured. Most
log entries are not set up to be understood easily by humans, with some
entries consisting of numeric values while others include cryptic text,
hexadecimal codes, or error codes. The analysis of this data and finding
correlations faces two main difficulties: first, the volume of RAS logs
makes the manual inspection difficult; and second, the unstructured nature
and idiosyncratic properties of log data produced by each subsystem log 
adds another dimension of difficulty in identifying implicit correlation 
among the events recorded. Consequently, the usage of log data is, in practice,
largely limited to detection of mere occurrences of known text patterns
that are already known to be associated with certain types of events.

As the scale and complexity of HPC systems continues to grow,
the storage, retrieval, and comprehensive analysis of the log data is a
significant challenge. In future extreme scale HPC systems the 
massive volume of monitoring and log data makes manual inspection and 
analysis impractical, and therefore poses a data analytics challenge. 
To address this challenge, scalable methods for processing log and 
monitoring data are needed. This will require storing the enormous data 
sets in flexible schemas based on scalable and highly available database 
technologies, which can support large-scale analytics with low latencies,
and high performance distributed data processing frameworks to support
batch, real-time, and advanced analytics on the system data. 

In this paper, we introduce a scalable HPC system data analytics
framework, which is designed to provide system log data analysis
capabilities to a wide range of researchers and engineers including system
administrators, system researchers, and end users. The framework
leverages Cassandra, a NoSQL distributed database to realize a scalable and
fast-response backend for high throughput read/write operations, the Apache 
Spark for supporting rapid analysis on the voluminous system data. The 
framework provides a web-based graphical, interactive frontend interface
that enables users to track system activity and performance and visualize 
the data. Using the framework, users can navigate spatio-temporal event space 
that overlaps with particular system events, faults, application runs, and 
resource usage to monitor the system, extract statistical features, and identify
persistent behavioral patterns. End users can also visually inspect trends
among the system events and contention on shared resources that occur during the run 
of their applications. Through such analysis, the users may find sources of 
performance anomalies and gain deeper insights into the impact of various system
behaviors on application performance. 

The rest of the document is organized as follows: Section
\ref{sec:DataModel} presents the data model and the design considerations that 
influenced the architecture of our framework. Section \ref{sec:Architecture} 
details the architecture of our framework and how it has been adapted to analyze 
data from the Titan supercomputer at the Oak Ridge Leadership Computing Facility 
(OLCF). Section \ref{sec:RelatedWork} surveys related works in HPC monitoring 
frameworks and the analysis of log data. Finally, Section \ref{sec:Conclusion} 
concludes the paper with a discussion on potential future directions.

\begin{figure}[!t]
\centering
\includegraphics[height=5.2cm]{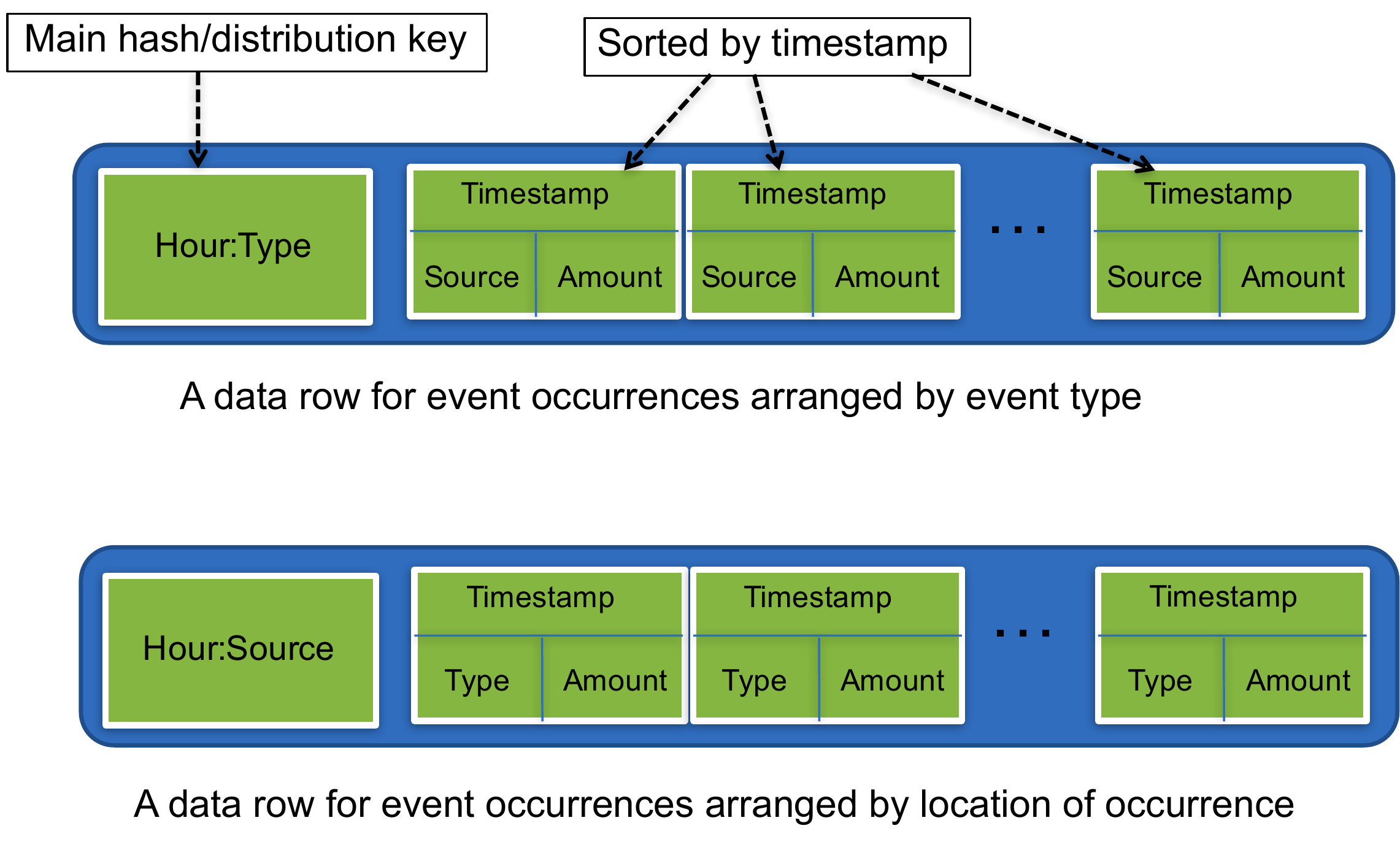}
\caption{Schemas for event occurrences: event schema ordered by time of occurrence (Top) and by location of occurrence (Bottom)}
\label{fig:event_schema}
\end{figure}

\begin{figure}[!t]
\centering
\includegraphics[height=5.2cm]{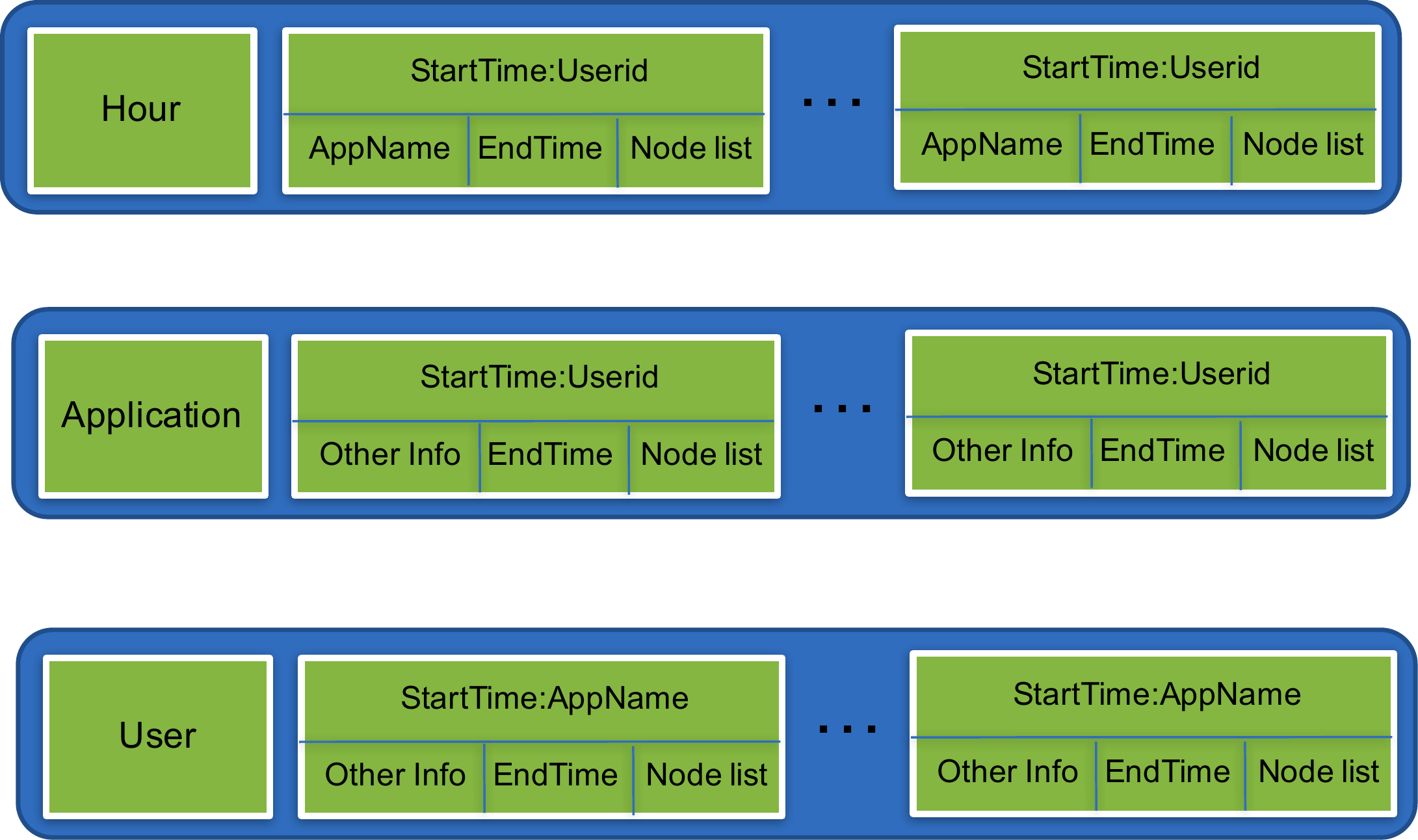}
\caption{Schemas for application runs: Application schema ordered by time of occurrence (Top), by name of application (Middle), and by users (Bottom)}
\label{fig:application_schema}
\end{figure}

\section{Data Model}
\label{sec:DataModel}

The monitoring infrastructure in supercomputing systems produces data 
streams from various sensors, which captures the resource and capacity 
utilization, power consumption, cooling systems, application performance, 
as well as various types of faults, errors and failures in the system. 
With the rapid increase in the complexity of supercomputing systems due 
to the use of millions of cores, complex memory hierarchies, and 
communication and file systems, massive amounts of monitoring data must 
be handled and analyzed in order to understand the characteristics of 
these systems and correlations between the various system measurements. 
The analysis of system monitoring data requires capturing relevant sensor 
data and system events, storing them in databases, developing analytic 
models to understand the spatial and temporal features of the data, or 
correlation between the various data streams, and providing tools capable 
of visualizing these system characteristics, or even building predictive 
models to improve the system operation. With the explosion in the monitoring 
data, this rapidly becomes a big data challenge. Therefore, to handle the 
massive amounts of system monitoring data and to support capabilities for 
more rigorous forms of user defined analytics, we adopt storage solutions 
designed to handle large amounts of data, and an in-memory data processing 
framework.

\subsection{Design Considerations}
An implementation of an HPC system log analytics framework should start
with extracting, transforming, and loading (ETL) of log data into a
manageable database that can serve a flexible and rich set of queries
over large amount of data. Due to the variety and volume of the 
data, we considered flexibility and fast performance to be two key 
design objectives of the framework. For an analytics framework to be  
successfully used for current and emerging system architectures, we
placed emphasis on the following design considerations for the backend 
data model:

\begin{itemize}
\item {\bf Scalability}: The framework needs to store historical 
  log data as well as future events from the monitoring frameworks. 
  The data model should be scalable to accommodate an ever increasing 
  volume of data.
\item {\bf Low latency}: The framework work also needs to serve
  interactive analytics that require near-real time query responses
  for timely visual updates. The backend data model should operate
  with minimal latency.
\item {\bf Flexibility}: A single data representation, or schema, for
  the various types of events from different system components is not 
  feasible. The data model should offer flexible mechanism to add new 
  event types and modify existing schemas to accommodate changes in 
  system configuration, software updates, etc.
\item {\bf Time series friendly}: The most common type of log analytics 
  that are of interest to HPC practitioners are expected to be based on 
  time series data, which provide insights about the system's behavior
  over a user specified window of time.
\end{itemize}

We believe that these features will enable users to identify patterns 
among the event occurrences over time and explain the abnormal behavior 
of systems and the impact on applications. The foundation of the 
analytics framework on such a data model will support a variety of 
statistical or data mining techniques, such as association 
rules \cite{Piatetsky-Shapiro:1991}, decision trees \cite{Quinlan:1986}, 
cross correlation \cite{Bansal:2004}, Bayesian network 
\cite{Heckerman:1997}, etc., to be applied to the system log data. 

For supporting a broad range of analytics, the retention of the raw data in a 
semi-structured format will be greatly beneficial. However, we found the 
conventional relational databases (RDBMS) do not satisfy our requirements. 
First, a schema of a relational database, once created, is very difficult to 
modify, whereas the format of HPC logs tend to change periodically. Second, 
due to its support for the atomicity, consistency, isolation, and durability 
(ACID) properties and two-phase commit protocols, it does not scale. 
After investigating various database technologies, we found the Apache 
Cassandra \cite{Cassandra:Website} to be most suitable for building the 
backend data model for our design of log analytics framework. Cassandra, 
based on Amazon's Dynamo and Google's BigTable, is a column-oriented distributed 
database offering highly available (HA) services with no single point of failure. 
Cassandra, a hashing-based distributed database system, stores data in
\emph{tables}. A data unit of a table, also known as \emph{partition},
is associated with a hash key and mapped to one or more nodes of a
Cassandra cluster. A partition can be considered as a data row that
can contain multiple column families, where each family can be of a
different format. 
Cassandra's performance, resiliency, and scalability come
from its master-less ring design, which unlike a legacy master-slave
architecture gives an identical role to each node. With a
replication option that is implemented on commodity hardware,
Cassandra offers a fault tolerant data service. Also with its column
oriented features, Cassandra is naturally suitable for handling data
in sequence, regardless of data sizes. When data is written to
Cassandra, each data record is sorted and written sequentially to
disk. When a database is queried, data is retrieved by row key and
range within a row, which guarantees a fast and efficient search.

\subsection{Data Model Design}
Our data model is designed to initially study the operational behavior of 
the Titan supercomputer hosted by the Oak Ridge National Laboratory. 
The framework is designed to study Titan's system logs collected
from console, application and network logs, which contain timestamped 
entries of critical system events. The data model is designed to 
capture various system events including, machine check exceptions, 
memory errors, GPU failures, GPU memory errors, Lustre file system 
errors, data virtualization service errors, network errors, application 
aborts, kernel panics, etc.  
 
We have created a total of eight tables to model system information, 
the types of event we monitor, occurrences of events, and
application runs. The partitions for events are designed to disperse
overheads in both reading and writing data evenly over to the cluster
nodes. Fig~\ref{fig:event_schema} shows how a partition is mapped
to one of the four nodes by its hash key of hour and type combination.

\begin{itemize}
\item \emph{nodeinfos}
\item \emph{eventtypes}
\item \emph{eventsynopsis}
\item \emph{event\_by\_time}
\item \emph{event\_by\_location}
\item \emph{application\_by\_time}
\item \emph{application\_by\_user}
\item \emph{application\_by\_location}
\end{itemize}

The \emph{nodeinfos} contains information about the system including the
position of a rack (or cabinet) in terms of row and column number, the
position of a compute node in terms of rack, chassis, blade, and
module number, network and routing information, etc. 
Each node in the Titan system consists of a AMD CPU and a NVIDIA GPU. 
Each CPU is a 16-core AMD Opteron 6274 processor with 32 GB of DDR3 memory 
and each GPU is a NVIDIA K20X Kepler architecture-based GPU with 6 GB of 
GDDR5 memory. The system uses Cray Gemini routers, which are shared 
between a pair of nodes.  
Each blade/slot Titan supercomputer consists of four nodes. Each cage 
has eight such blades and a cabinet contains three such cages. The 
complete system consists of 200 cabinets that are organized in a grid of 
25 rows and 8 columns. The \emph{nodeinfo} enables spatial correlation and 
analysis of events in the system.

The two tables \emph{event\_by\_time} and \emph{event\_by\_location} 
store system event information from two perspectives, time and location 
to facilitate spatio-temporal analysis. An event in our data model is 
defined as occurrence(s) of a certain type reported at a particular 
timestamp. An event is also associated with the location (or the source 
component) where it is reported. The two tables illustrate these dual 
representations of an event as illustrated in Fig~\ref{fig:event_schema}. 
The first table structure associates an event with its type and the hour 
of its occurrence; all events of a certain type generated at a certain hour
are stored in the same partition. In contrast, the second table
structure associates an event with hour and location; all events,
regardless of their type, generated at a certain hour for the same
component, are stored in the same partition. Note that each partition
stores events sorted by their timestamps, which is a time series
representation of events that is one hour long. This facilitates to
support a spatio-temporal query.

For data about user application runs, we added another dimension:
users.  More specifically, three tables to represent user application
runs from perspectives of time, application, and user (see
Fig~\ref{fig:application_schema}). Readers can find this as a set
of denormalized views on application runs. Note however, although all
application runs in each partition type are depicted the same, in
fact, each application run may include columns unique to it. For
example, a column named as \emph{Other Info} may include multiple
sub-columns to represent different information.

\section{Architecture of the Log Analytics Framework}
\label{sec:Architecture}

\begin{figure}[!t]
\centering
\includegraphics[height=7.0cm, width=\linewidth]{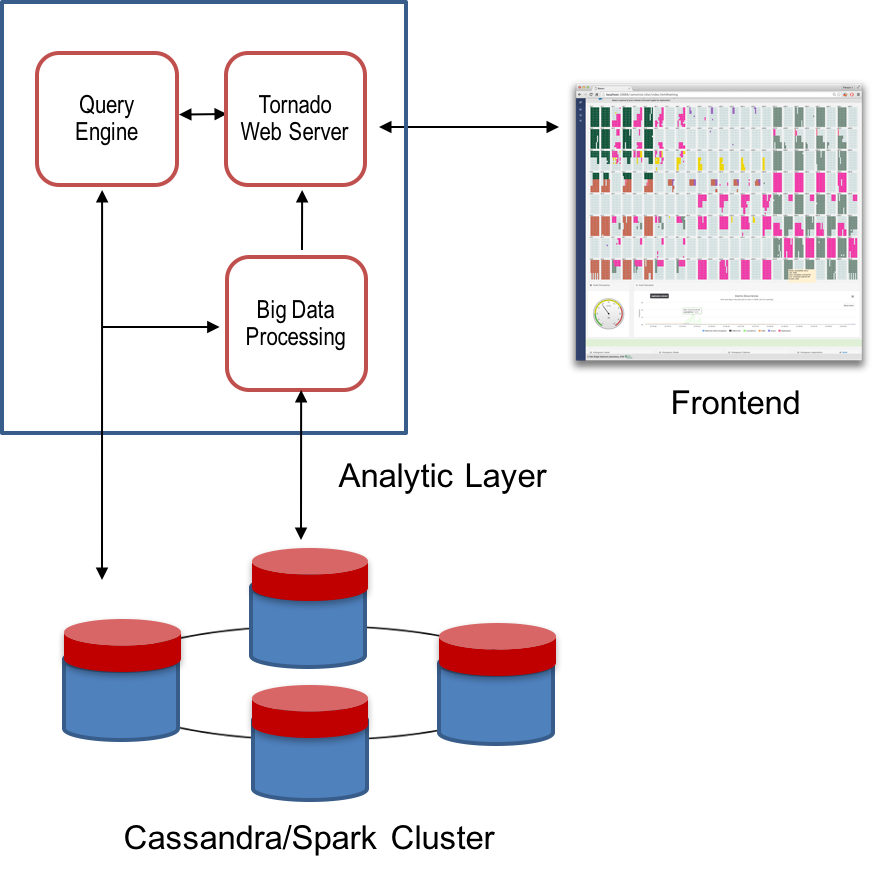}
\caption{Overall architecture of the Log Analytics Framework
  consisting of the Cassandra distributed NoSQL database and the
  Apache Spark in-memory data processing engine}
\label{fig:3-tier}
\end{figure}

The layout of the overall architecture, illustrated in
Fig~\ref{fig:3-tier}, consists of three main components:
a web-based frontend, the data analytics server, and the backend
distributed database. The frontend, which consists of a client-side
application, adopts a web interface allowing users to create queries
for the analysis of the data as well as for visual inspection of
log data and application runs in both spatial and temporal domain.
The analytics server translates data query requests received from the
frontend and relays them to the backend database server in the form
of Cassandra Query Language (CQL) queries. The query results
from the backend are returned to the analytics server either as
data that may be transmitted to the frontend, or as intermediate data
for further processing. The backend distributed NoSQL database
stores and manages Titan system logs and user application logs. The
backend server communicates with the analytics layer through a RESTful
interface. Query results are sent in JSON object format to avoid data
format conversion at the frontend. This framework is currently being
deployed at the ORNL's Compute and Data Environment for Science (CADES),
which provides flexible computing, data storage and analytics
infrastructure.

\subsection{Analytic Server and Backend Database}
The analytics server consists of a web server, a query processing engine,
and a big data processing engine. The user queries are received by the web
server, translated by the query engine, and either forwarded to the
backend database, or the big data processing unit depending on the type
of a user query. Simple queries are directly handled by the query
engine, and complex queries are passed to the big data processing unit.
The big data processing unit initiates a Spark session over the Spark
cluster that reside on the same nodes with Cassandra.

Since the analytic framework intends to serve numerous users, who
may require long-lived connections and may expect delayed responses from
the server for non-trivial analytics, we chose the Tornado framework 
\cite{Tornado}, which provides a web server and asynchronous networking 
library. Tornado supports non-blocking I/O, which makes it suitable
for \emph{long-polling} and \emph{WebSockets} to implement the long-living
connections from the frontend web-based client.

\begin{figure}[!t]
\centering
\includegraphics[height=5.2cm, width=\linewidth]{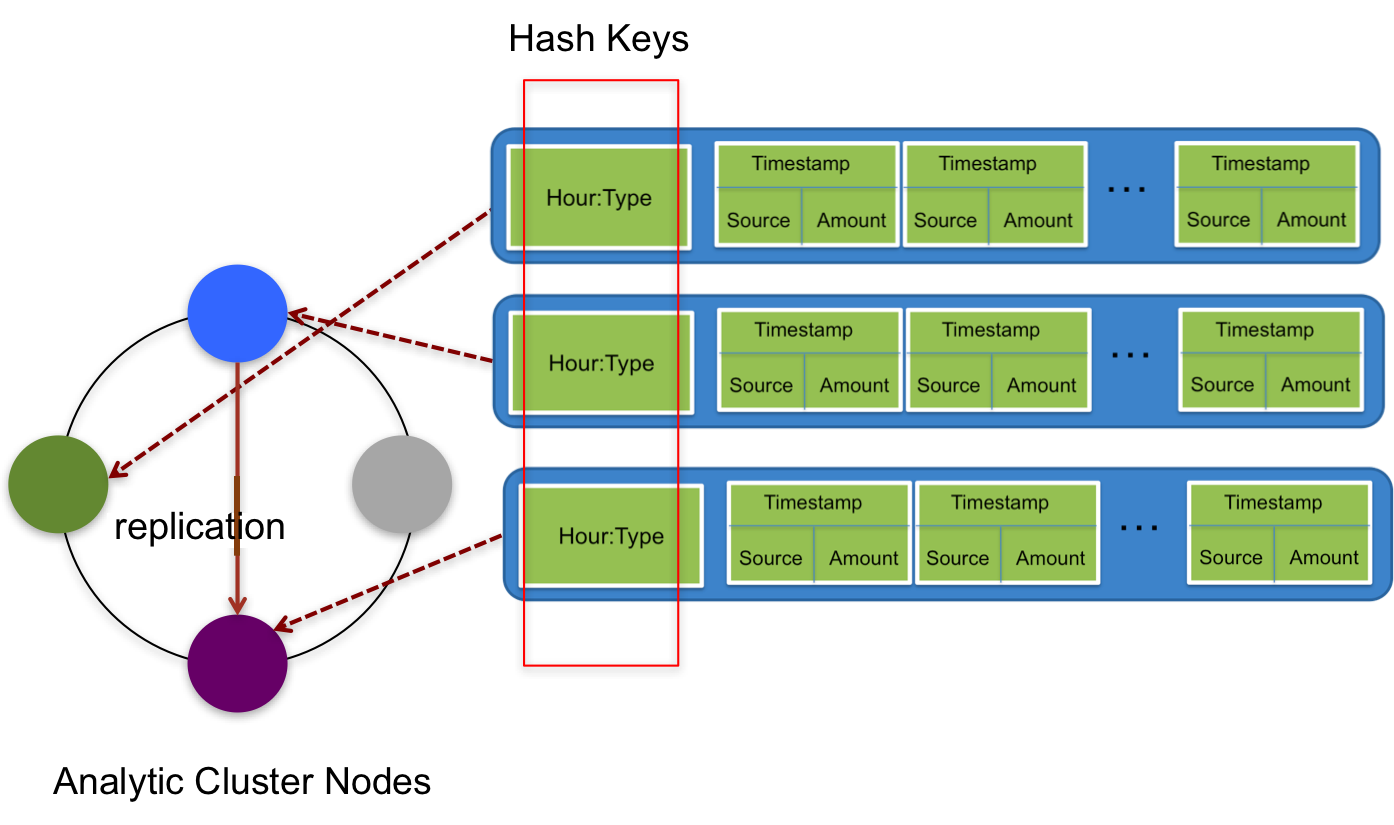}
\caption{Event partitions mapped to Cassandra nodes by hour and event types}
\label{fig:partitions}
\end{figure}

The Cassandra backend database server and the Apache Spark cluster are
installed over the same 32 virtual machine (VM) instances in CADES, that is,
a pair of a Spark worker node and a Cassandra node runs together in each of
the 32 VMs. We selected this configuration to maximize data
locality for the computation performed by the analytic algorithms of the big 
data processing unit. As described in Section~\ref{sec:DataModel}, a 
\emph{partition} of a table is defined by a combination of hour, user, application,
location, and event type representing the data from a specific view
(this will be defined as a \emph{context} below). Each table is
distributed over the entire cluster retaining time ordered data
entries within each partition. The big data processing unit consists of
a set of Spark computations that perform MapReduce operations over
time ordered data spread across the cluster by a context. By associating local
partitions with the same local Spark worker, the big data processing
unit performs analytics efficiently. Fig~\ref{fig:partitions} illustrates
an example of partitions for event occurrences that are mapped to nodes
on the basis of the event hour and event type.

\begin{figure}[!t]
\centering
\includegraphics[height=7.5cm, width=\linewidth]{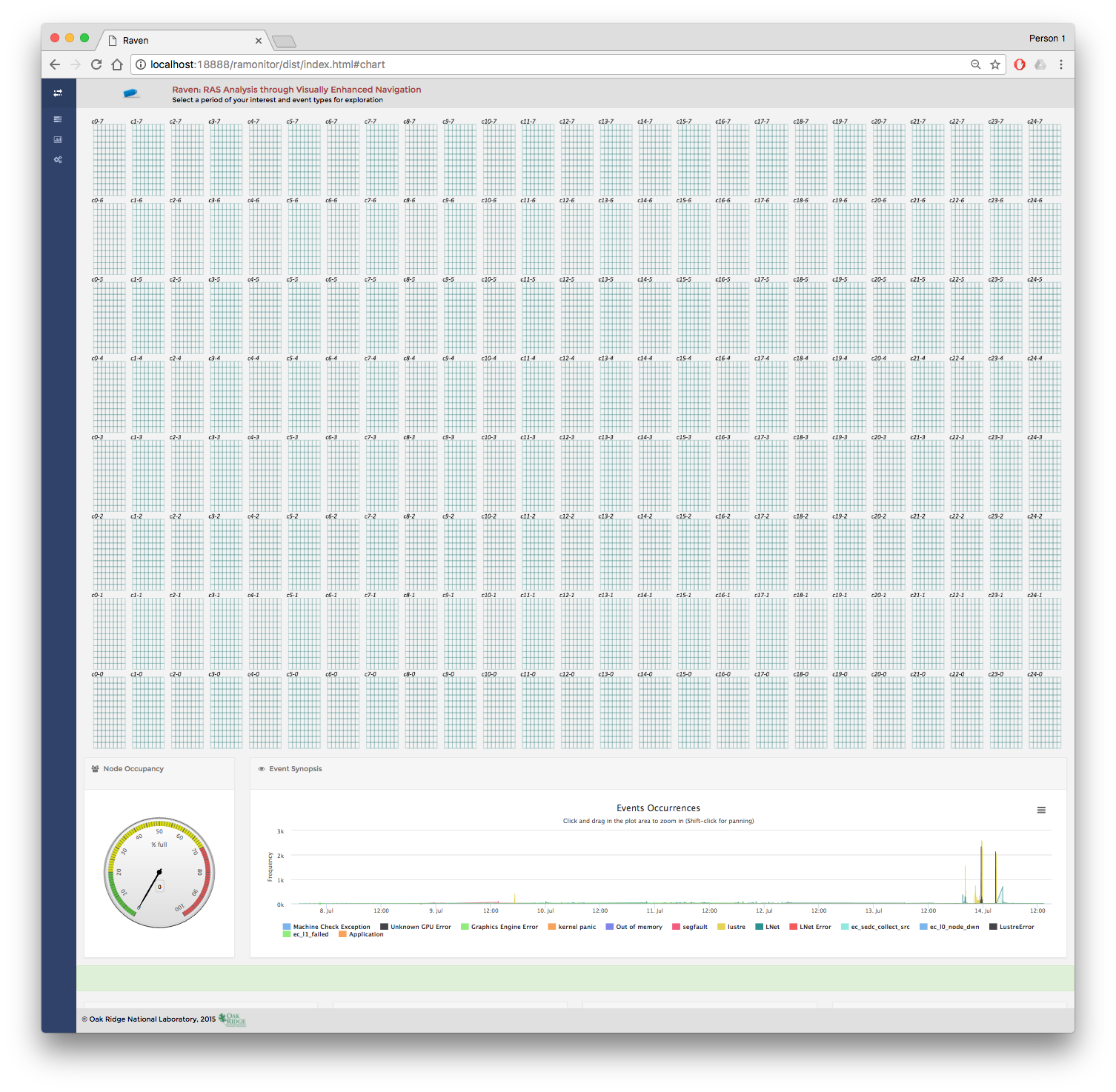}
\includegraphics[height=7.5cm, width=\linewidth]{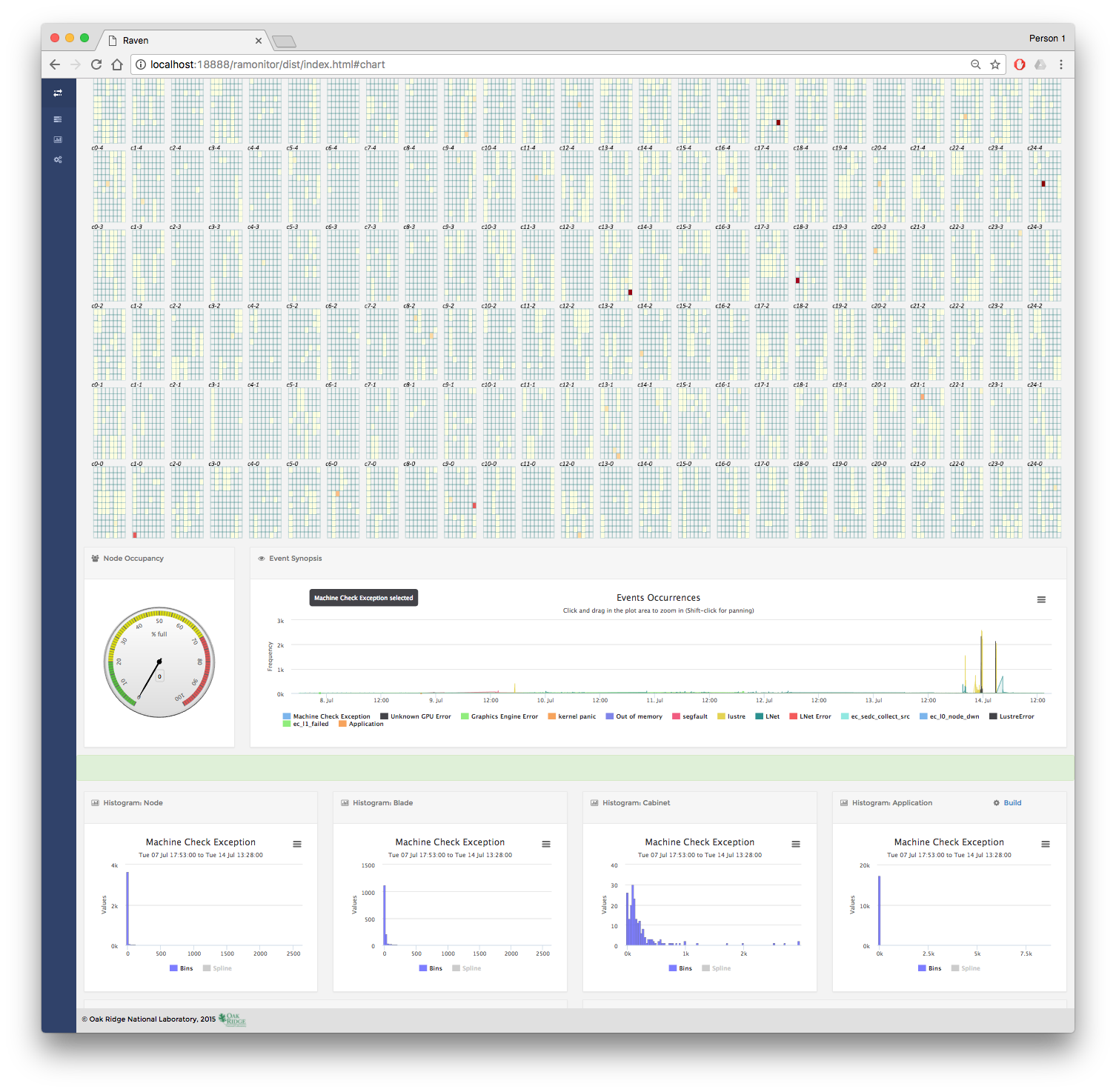}
\caption{The physical system map and the temporal map of the frontend (Top); 
  Distribution of an event type over a selected period as a heat map on the physical system map
  and event histograms
  (Bottom).}
\label{fig:heatmap}
\end{figure}

\subsection{Frontend: Client Module}
The frontend provides a window to the system log data. Users interact 
with the frontend to inspect the system status or perform analytics on 
log data. Every interaction with the frontend is translated into a query 
in Javascript Object Notation (JSON) format and delivered to the analytic 
server. The current frontend provides visualization of events and application
runs in both spatial and temporal dimension on physical system map and
time interval map.  The visualization is implemented using D3
\cite{Bostock:2011} package and HTML5 canvas.

Users interact with the framework by creating a $context$.  A context
is selected on the basis of event type, application, location, user, time
period, or a combination of these, over which the system status is
defined and examined. By selecting a context, important
insights about the system status can be extracted. The selection of an
appropriate context also helps in identifying the root cause of failure 
events. When a context is created the appropriate query is passed to 
the data analytic server to retrieve data. The frontend allows users to 
choose desired contexts and results by interacting with:

\begin{itemize}
\item {The physical system map}
\item {The temporal map}
\item {The event types map}
\item {The user/application map}
\item {The tabular map of raw log entries}
\end{itemize}

\begin{figure}[!t]
\centering
\includegraphics[height=7.5cm, width=8.8cm]{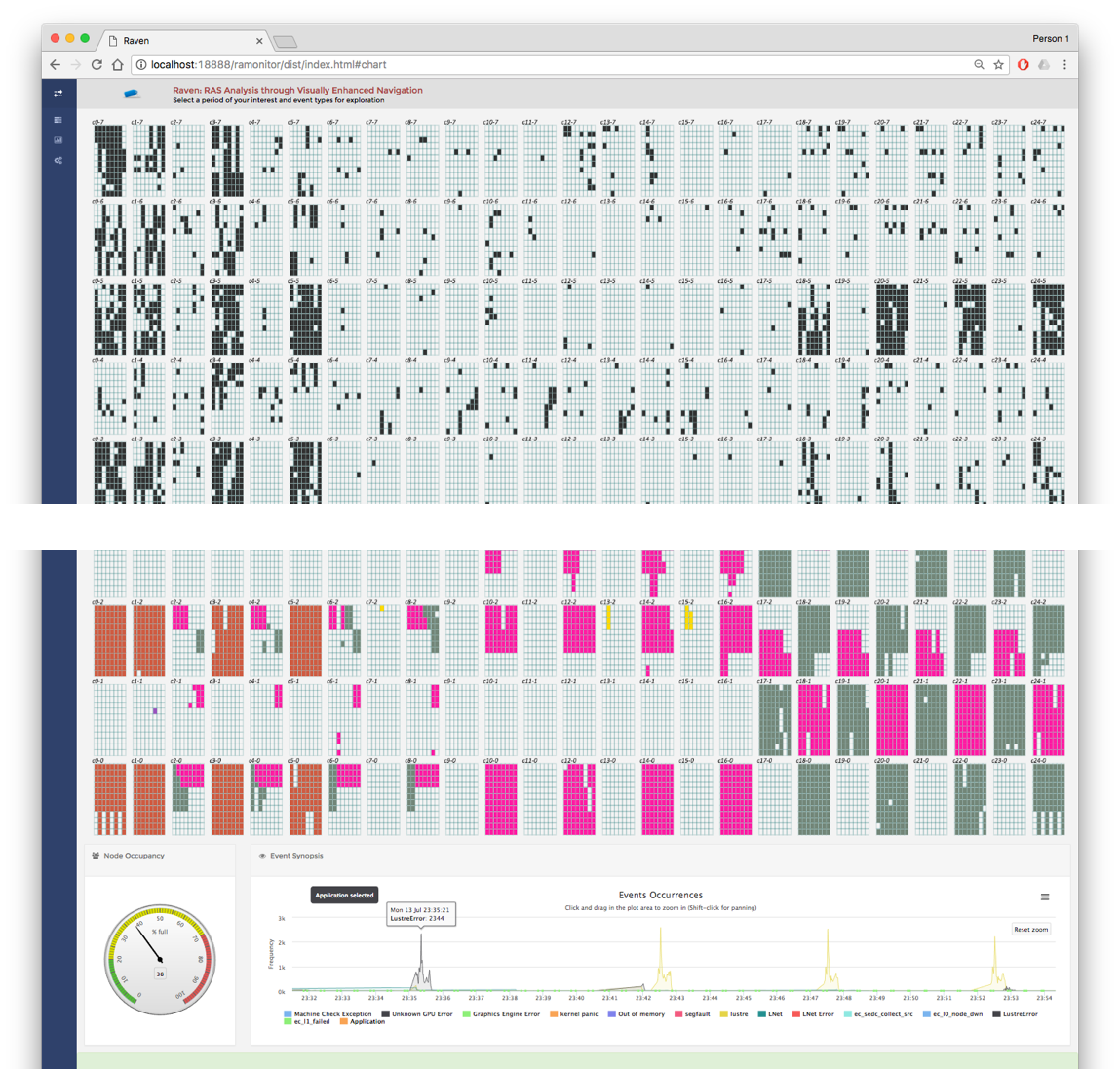}
\caption{Event occurrences (Top) and Application placement (Bottom) 
         rendered on the Physical System Map}
\label{fig:error_apps}
\end{figure}

While the physical system map shows the spatial placement of racks
(or cabinets) and the individual nodes within each rack, the temporal
map shows occurrences of events over a time interval.
Fig~\ref{fig:heatmap}-(Top) shows the physical system map and the
temporal map. Event occurrences, or application displacements, are
displayed on the physical system map.  Using the event type map and
the temporal map, users can select an event type of interest at a
particular time. The occurrences of the selected event type at the
specified timestamp are shown on the compute nodes where they occurred
in the physical system map.  Likewise, displacements of all
applications that were running at the time of selection, once users
select using the user/application map, are shown on the nodes in the
physical system map. Fig~\ref{fig:error_apps} shows Lustre error
occurrences on each compute node (Top) and the placement of user
applications (Bottom) at the specified timestamp. Users can also
select a hardware component such as compute node in the physical
system map. Thus, the physical system map, the temporal map, the event
types map, and the user/application map are essentially interactive
visualization components that allow discovery of correlations among
events, locations, and applications by tracing occurrences and
progression of events.

The temporal map represents a selected time interval. Users can
repeatedly select sub-intervals of interest for narrowed
investigations. With a selected interval, users can extract basic
statistics about event occurrences. First, users can create a heat map
representation of the occurrences of an event type within the interval
on the physical system map, which illustrates whether the event
occurrences were unusually higher (or lower) in some parts of the
system compared to the rest of the parts. In addition, users can also
get distributions of the event occurrences over cabinets, blades,
nodes, and applications. These two different types of view (heat map
and distributions) offer complementary insights on normal or abnormal
occurrences of a certain event type observed during a selected period.
Fig~\ref{fig:heatmap}-(Bottom) shows that Machine Check Exception
(MCE) errors occurred abnormally high in some compute nodes over a
selected time period.

\subsection{Big Data Processing using the Frontend}
The big data processing unit intends to serve a wide range of users for
intensive analytic processing. We are currently developing
log data analytic application program interfaces (APIs) through which
users can connect to the analytic server from their chosen
applications. The frontend also offers a set of basic analytics
capabilities utilizing big data processing unit.

\begin{figure}[!t]
\centering
\includegraphics[height=6.0cm, width=8.8cm]{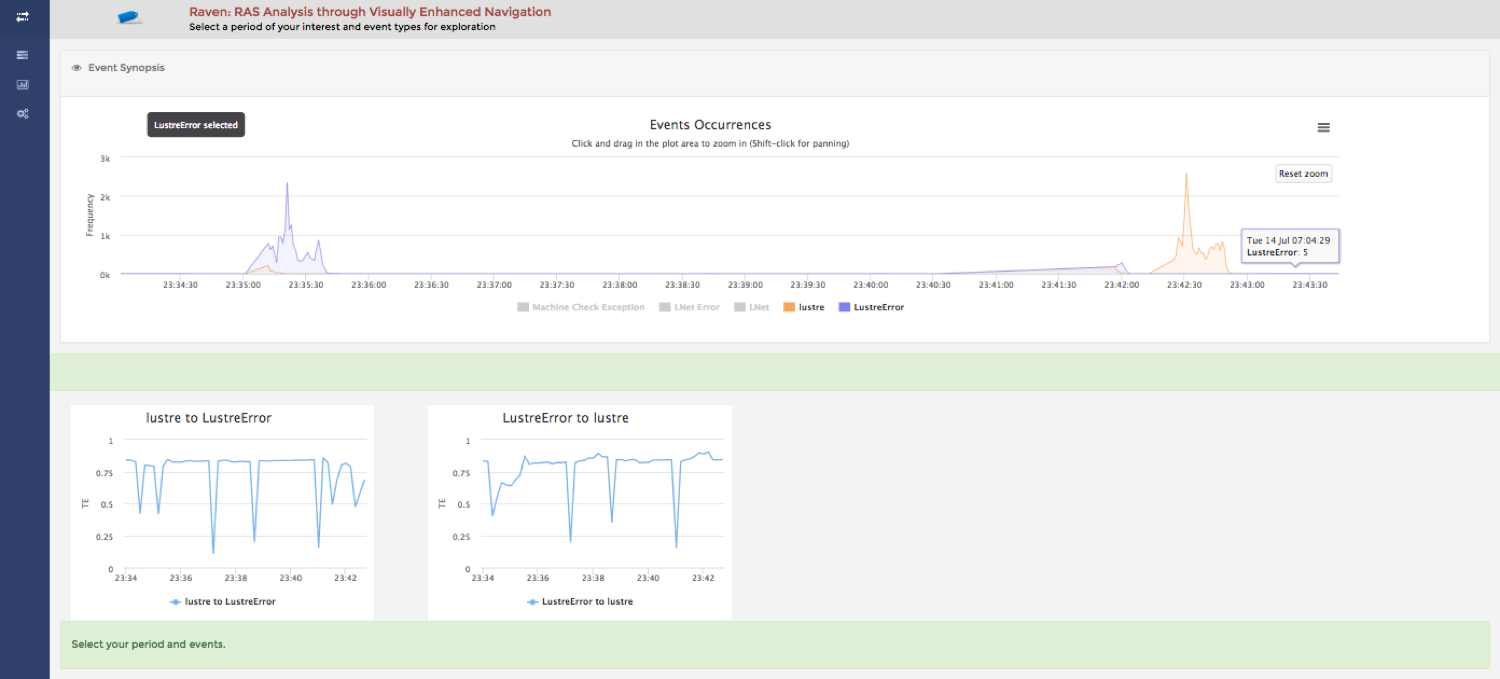}\\
\vspace{5mm}
\includegraphics[height=6.0cm, width=8.8cm]{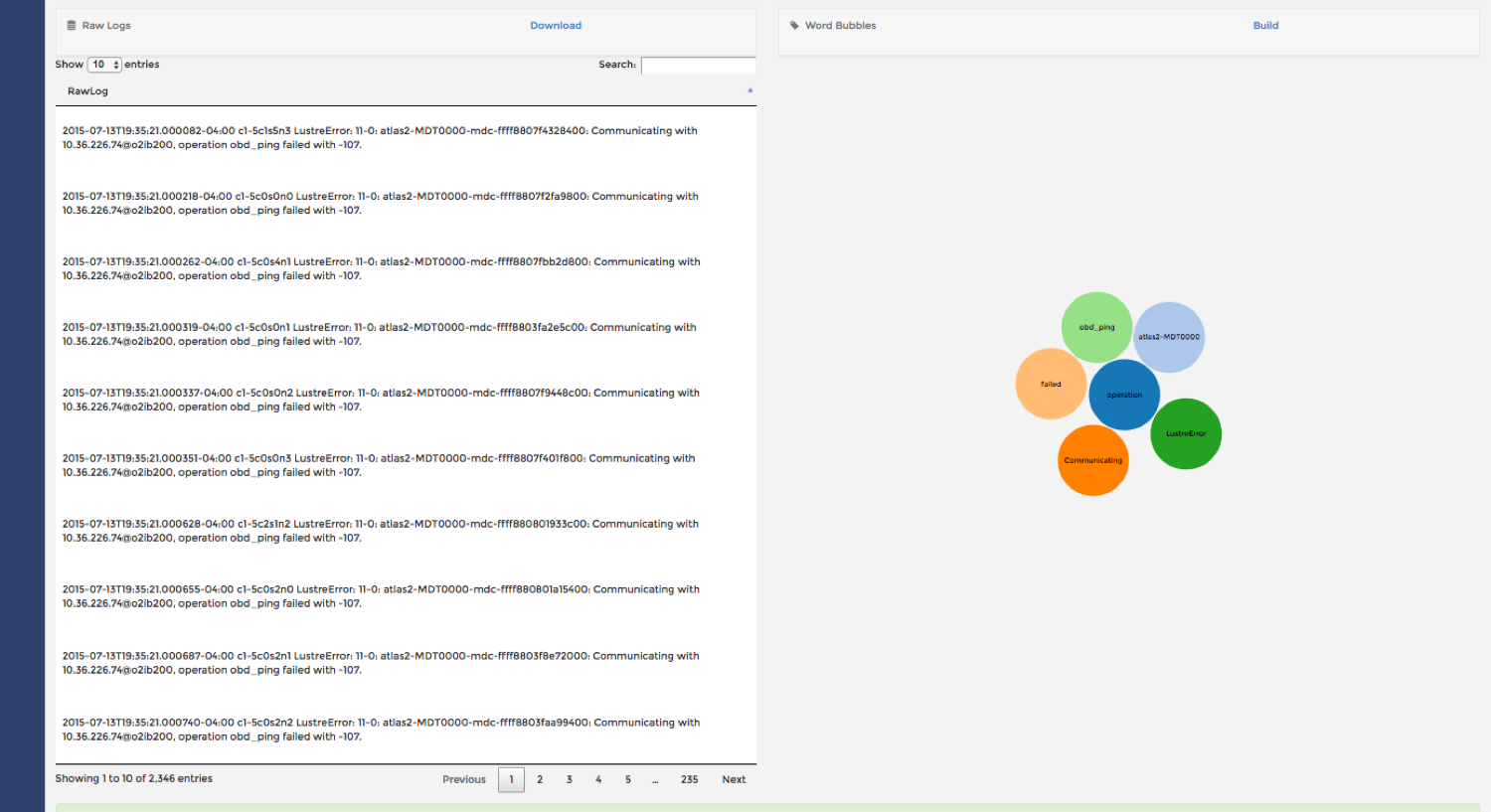}
\caption{Transfer Entropy plot of two event types measured within a
  selected time interval (Top). Raw log entries shown in the tabular
  map and importance words from logs illustrated as bubbles (Bottom).}
\label{fig:information}
\end{figure}

First, the heat map representation and various distributions of event
occurrences over a selected time interval, which are mentioned above,
are computed by the big data processing. Second, the investigation of
correlation between two event occurrences within a selected time
interval, which can provide a causal relationship between the two, is
also processed by the big data processing unit.
Fig~\ref{fig:information}-(Top) shows the transfer entropy plot of
two events measured within a selected time window.

Also, basic text analytics are supported by the big data processing unit.
Identification of $important$ keywords (either letters or alphanumeric
values) from raw system logs often helps understanding the system 
status given a massive number of system events logged. Examples include 
events from the filesystem, the network subsystem, etc. For example, 
Lustre log message contains descriptions regarding status of hardware, an I/O
transaction, the peers of the log generator, etc. These information
are written in texts, hexadecimal numbers, or special characters. Once
properly filtered, each Lustre event message can be transformed into a
set of words that represents the event occurrence as a point in a
metric space. Such transformations typically involve \emph{word counts}
and/or term frequency-inverse document frequency (TF-IDF) of
log messages. Note here a Lustre message is treated as a document
from a conventional text analysis point of view. The temporal event
view in Fig~\ref{fig:information}-(Bottom) shows a period when tens
of thousands Lustre error messages were generated. As shown in the
map, it was a system wide event that lasted several minutes afflicting
most of compute nodes and applications running therein.  In many
cases, the root causes of such a system-wide event are abnormal
behaviors of either hardware or a system software component of which
negative impacts propagate over to the entire system. Although it may
be a single source problem, it requires to sift through a large volume
of Lustre event logs to identify the problem components. We found that
a simple word counts, which is rapidly executed by Spark, can locate
the source of the problem.  Fig~\ref{fig:information}-(Bottom)
shows word bubbles as the result of text analysis on raw Lustre event
logs, which illustrates an object storage target is not responding.

\subsection{Data Ingestion}
The log analytics framework is designed to ingest new event data in
two different modes: batch import and real-time streaming. The batch
import is a traditional ETL procedure that involves 1) collocation of
all data, 2) parsing the data in search for known patterns for each
event type (typically defined as regular expressions), and 3) batch
upload into the backend database. The batch import is used when a
new event type is identified and all occurrences in the historical 
data must be collected. Since such an update may require huge 
computational overheads, the analytic framework implements parsing 
and uploading using Apache Spark.

The real-time streaming mode, which is currently being developed through
a collaboration with the high-performance computing operation group of the
Oak Ridge Leadership Computing Facility (OLCF), intends to
facilitate online analytics such as real time failure detection by
monitoring recent event streams. The OLCF is developing event
producers that not only parse real-time streams from log sources but
also publish each event occurrence from the streams.  Each event
occurrence is published to an Apache Kafka message bus that is available
to consumers subscribing to the corresponding topic.
For example, event logs of a Lustre filesystem are
generated at multiple places: the servers (OSSes, MDSes, and MGS), and
the clients at each compute node. In addition, lower level components
(e.g. disk controllers) also generate separate logs.

The OLCF has deployed Kafka on top of OpenShift Origin, which is a scalable
container-based framework for scheduling applications.  This deployment method
allows elastic scale-out of Kafka nodes and backend databases to accommodate large
upswings in datatype growth.  As new supercomputing resources and high performance
filesystems are brought online, additional OpenShift pods will be deployed as
necessary to handle additional log event types.  Conversely, as analytics
frameworks grow in size and complexity, the Kafka install will be scaled out
to meet the demand of any consumers.

To receive event streams from the Kafka instance, the analytic
framework places a subscriber that delivers event messages to
Spark streaming module that in turn converts and places all event
occurrences into the right partitions. Event occurrences of the same
type and same location are coalesced into a single event if they are
timestamped the same. For this, the time window of the Spark streaming
is set to one second. This real-time upload component will be further
extended to include various online analytic modules.

\section{Related Work}
\label{sec:RelatedWork}

Various monitoring frameworks are used in large-scale computing systems for understanding the use of system resources by applications, the impact of competition for shared resources and for the discovery of abnormal system conditions in the system. Tools such as Ganglia \cite{Massie:2004} and Nagios \cite{Barth:2008} are widely used for in HPC cluster and grid systems as well as in enterprise clusters. OVIS \cite{Brandt:2006} provides a suite of monitoring and analysis tools for HPC systems that provides finer grained monitoring to enable understanding platform resource utilization characteristics of applications. 

Several studies have sought to analyze failures in large-scale systems to characterize the reliability of the system. These studies attempt to characterize the root causes of single-node failures as well as system-wide failures from manual reports and system event logs \cite{Martino:2014}. These studies perform post-mortem analysis on the system logs to extract the statistical properties of system errors and failures \cite{Sahoo:2004} \cite{Schroeder:2010}. For the analysis of logs of large-scale systems, certain approaches apply filtering, followed by extraction and categorization of error events \cite{Liang:2005} \cite{Oliner:2007} \cite{Pecchia:2011}. Other analyses use approaches such as time coalescing \cite{Martino:2012}. Some studies have focused on analysis of failure characteristics of specific subsystems or system components in HPC systems, such as disks \cite{Schroeder:2009}, DRAM memory \cite{Schroeder:2009} \cite{Hwang:2012} \cite{Sridharan:2013}, graphical processing units (GPU) \cite{Tiwari:2015}. These studies sanitize the system logs using manual failure reports, or extract specific events of interest, to compute the relative failure frequencies for various root causes and their mean and standard deviation in contrast to our framework, which mines for insights from the unstructured raw data. Our approach is designed to handle massive amounts of heterogeneous monitoring and log data, which will be typical in future extreme-scale systems with complex hardware and software architectures. 

Based on the observation of characteristics of failure events and correlations between the events, models for failure prediction have been proposed \cite{Liang:2006} \cite{Gainaru:2012}. These prediction algorithms leverage the spatial and temporal correlation between historical failures, or trends of non-fatal events preceding failures to design remedial actions in the system's configuration, scheduling algorithms to mitigate the adverse impacts of failure events.

\section{Conclusion}
\label{sec:Conclusion}
With the ever-growing scale and complexity of high performance
computing (HPC) systems, characterizing system behavior has
become a significant challenge. The systems produce and log vast 
amounts of unstructured multi-dimensional data collected using a 
variety of monitoring tools. The tools and methods available today 
for processing this log data lack advanced data analytics 
capabilities, which makes it difficult to diagnose and completely 
understand the impact of system performance variations, fault and 
error events in the system on application performance. To handle
the massive amounts of system log data from a diverse
set of monitoring frameworks and rapidly identify problems and 
variations in system behavior, it is essential to have scalable tools 
to store and analyze the data.

In this paper, we introduced a scalable HPC log data analytics framework 
based on a distributed data and computation model. The framework defines
a time-series oriented data model for HPC log data. We leverage 
big data frameworks, including Cassandra, a highly scalable, 
high-performance column-oriented NoSQL distributed database, and Apache 
Spark, a real-time distributed in-memory analytics engine.
We presented a data model designed to facilitate log data analytics
for system administrators and researchers as well as end users who
are often oblivious to the impact of variations and fault events on 
their application jobs.  

Our log analytic framework has been tested with Titan supercomputer 
at the Oak Ridge Leadership Computing Facility's (OLCF). Although the
framework is still evolving, with new analytics modules being
currently developed, the preliminary assessment shows that the framework 
can provide deeper insights about the root causes of system faults, and
abnormal behaviors of user applications. It also enables statistical
analysis of event occurrences and their correlations on a spatial and 
temporal basis. These capabilities will be valuable when deploying a new 
HPC system in the pre-production phase, as well as during operational 
lifetime for fine tuning the system.  

While our existing framework improves upon the state-of-the-art in HPC 
log data processing, there is much room to improve. As future work, 
we are planning several enhancements and improvements to the framework. 
First, new and composite event types will need to be
defined for capturing the complete status of the system. This will involve
event mining techniques rather than text pattern matching. Second, the
framework will need to develop application profiles in terms of event
occurred during its runs. This will help understand correlations 
between application runtime characteristics and variations observed in 
the system on account of faults and errors. Finally, the framework will 
need to support advanced statistical techniques, incorporate machine learning
algorithms, and graph analytics for more comprehensive investigation of log
 and monitoring data. 


\section*{Acknowledgment}
This material is based upon work supported by the U.S. Department of
Energy, Office of Science, Office of Advanced Scientific Computing
Research, program manager Lucy Nowell, under contract number
DE-AC05-00OR22725. This work was supported by the Compute and Data
Environment for Science (CADES) facility and the Oak Ridge
Leadership Computing Facility at the Oak Ridge National Laboratory,
which is managed by UT Battelle, LLC for the U.S. DOE (under the
contract No. DE-AC05-00OR22725).
\bibliographystyle{IEEEtran}
\bibliography{IEEEabrv,references}

\end{document}